\documentstyle{aipproc}

\newcommand{\ud}{\mathrm{d}}
\begin{document}
\title{Variational Approach in Wavelet Framework to Polynomial
Approximations of Nonlinear Accelerator Problems}
\author{A.~Fedorova$^*$,  M.~Zeitlin$^*$ and Z.~Parsa$^\dagger$}
\address{$^*$Institute of Problems of Mechanical Engineering,
 Russian Academy of Sciences, 199178, Russia, St.~Petersburg,
  V.O., Bolshoj pr., 61, e-mail: zeitlin@math.ipme.ru and\\
 $^\dagger$Dept. of Physics, Bldg.~901A, Brookhaven National Laboratory,
Upton, NY 11973-5000, USA}
\maketitle

\begin{abstract}
In this paper we
present applications of methods from
wavelet analysis to polynomial approximations for
a number of accelerator physics problems.
According to variational approach
in the general case we have the solution as
a multiresolution (multiscales) expansion in the base of compactly
supported wavelet basis.
We give extension of our  results to the cases of periodic orbital
particle motion and arbitrary variable coefficients.
Then we consider more flexible variational
method which is based on biorthogonal wavelet approach.
Also we consider different variational approach, which is
applied to each scale.
\end{abstract}
\section{INTRODUCTION}
This is the first part of our two presentation in which we consider
applications of methods from wavelet analysis to nonlinear accelerator
physics problems. This is a continuation of  results from [1]-[6],
which is based on our approach  to investigation
of nonlinear problems -- general, with additional structures (Hamiltonian,
symplectic or quasicomplex), chaotic, quasiclassical, quantum, which are
considered in the framework of local (nonlinear) Fourier analysis, or wavelet
analysis. Wavelet analysis is a relatively novel set of mathematical
methods, which gives us a possibility to work with well-localized bases in
functional spaces and with the general type of operators (differential,
integral, pseudodifferential) in such bases.

We consider application of multiresolution representation to
general nonlinear dynamical system with polynomial type of nonlinearities.
In part II we consider this very useful approximation in
the cases of orbital motion in storage rings, a particle in the
multipolar field, effects of insertion devices on beam dynamics,
spin orbital motion.
Starting  in part III A from  variational formulation of
initial dynamical problem we
construct via multiresolution analysis (part III B)
 explicit representation for all dynamical variables in the base of
compactly supported (Daubechies) wavelets. Our solutions (part III C)
are parametrized
by solutions of a number of reduced algebraical problems one from which
is nonlinear with the same degree of nonlinearity and the rest  are
the linear problems which correspond to particular
method of calculation of scalar products of functions from wavelet bases
and their derivatives. Then  we consider further extension of our
previous results.  In part V we consider modification of our
construction to the periodic case,
in part VI we consider generalization of our approach
 to variational formulation in the biorthogonal bases of compactly
supported wavelets and in part VII to the case of variable coefficients.
In part IV we consider the different variational approach
which is based on ideas of para-products (A) and approximation
for multiresolution approach, which gives  us possibility
for computations in each scale separately (B).

\section{Problems and approximations}
Below we consider a number of examples of nonlinear accelerator physics
problems which are from the formal mathematical point of view
not more than nonlinear differential equations with polynomial
nonlinearities and variable coefficients.
\subsection{Orbital Motion in Storage Rings}

We consider as the main example the particle motion in
storage rings in standard approach, which is based on
consideration in [7].
Starting from Hamiltonian, which described classical dynamics in
storage rings
\begin{eqnarray}
{\cal H}(\vec{r},\vec{P},t)=c\{\pi^2+m_0^2c^2\}^{1/2}+e\phi
\end{eqnarray}
and using Serret--Frenet parametrization, we have the following
Hamiltonian for orbital motion in machine coordinates:

\begin{eqnarray}
{\cal H}(x,p_x,&z&,p_z,\sigma,p_\sigma;s)=
   p_\sigma-[1+f(p_\sigma]\cdot[1+K_x\cdot x+K_z\cdot z]\times\\ \nonumber
& &\Bigg\{ 1-\frac{[p_x+H\cdot z]^2 + [p_z-H\cdot x]^2}
{[1+f(p_\sigma)]^2}\Bigg\}^{1/2}\\\nonumber
& &+\frac{1}{2}\cdot[1+K_x\cdot x+K_z\cdot z]^2-\frac{1}{2}\cdot g\cdot
   (z^2-x^2)-N\cdot xz\\ \nonumber
& &+\frac{\lambda}{6}\cdot(x^3-3xz^2)
   +\frac{\mu}{24}\cdot(z^4-6x^2z^2+x^4)\\ \nonumber
& &+\frac{1}{\beta_0^2}\cdot\frac{L}{2\pi\cdot h}\cdot\frac{eV(s)}{E_0}\cdot
\cos\left[h\cdot\frac{2\pi}{L}\cdot\sigma+\varphi\right]\nonumber
\end{eqnarray}

Then, after standard manipulations with truncation of
power series expansion of square root we arrive to the following
approximated Hamiltonian for particle motion:

\begin{eqnarray}
{\cal H}&=&
   \frac{1}{2}\cdot\frac{[p_x+H\cdot z]^2 + [p_z-H\cdot x]^2}
{[1+f(p_\sigma)]}+
p_\sigma-[1+K_x\cdot x+K_z\cdot z]\\\nonumber&&\cdot f(p_\sigma)+
\frac{1}{2}\cdot[K_x^2+g]\cdot x^2+\frac{1}{2}\cdot[K_z^2-g]\cdot z^2-
  N\cdot xz+\\ \nonumber
& &\frac{\lambda}{6}\cdot(x^3-3xz^2)+\frac{\mu}{24}\cdot(z^4-6x^2z^2+x^4)\\ \nonumber
& &+\frac{1}{\beta_0^2}\cdot\frac{L}{2\pi\cdot h}\cdot\frac{eV(s)}{E_0}\cdot
\cos\left[h\cdot\frac{2\pi}{L}\cdot\sigma+\varphi\right]\nonumber
\end{eqnarray}

and the corresponding equations of motion:

\begin{eqnarray}
\frac{\ud}{\ud s}x&=&\frac{\partial{\cal H}}{\partial p_x}=
     \frac{p_x+H\cdot z}{[1+f(p_\sigma)]}; \nonumber\\
\frac{\ud}{\ud s}p_x&=&-\frac{\partial{\cal H}}{\partial x}=
\frac{[p_z-H\cdot x]}{[1+f(p_\sigma)]}\cdot H -[K^2_x+g]\cdot x+N\cdot z+
 \nonumber\\
  & & K_x\cdot f(p_\sigma)-\frac{\lambda}{2}\cdot(x^2-z^2)-
  \frac{\mu}{6}(x^3-3xz^2);\\
\frac{\ud}{\ud s}z&=&\frac{\partial{\cal H}}{\partial p_z}=
 \frac{p_z-H\cdot x}{[1+f(p_\sigma)]}; \nonumber\\
\frac{\ud}{\ud s}p_z&=&-\frac{\partial{\cal H}}{\partial z}=
     -\frac{[p_x+H\cdot z]}{[1+f(p_\sigma)]}\cdot H -
     [K^2_z-g]\cdot z+N\cdot x+ \nonumber\\
  & & K_z\cdot f(p_\sigma)-\lambda\cdot xz-
  \frac{\mu}{6}(z^3-3x^2z);\nonumber\\
\frac{\ud}{\ud s}\sigma&=&\frac{\partial{\cal H}}{\partial p_\sigma}=
 1-[1+K_x\cdot x+K_z\cdot z]\cdot f^\prime(p_\sigma)- \nonumber\\
& &\frac{1}{2}\cdot\frac{[p_x+H\cdot z]^2+[p_z-H\cdot x]^2}{[1+f(p_\sigma)]^2}
 \cdot f^\prime(p_\sigma)\nonumber\\
\frac{\ud}{\ud s}p_\sigma&=&-\frac{\partial{\cal H}}{\partial\sigma}=
\frac{1}{\beta_0^2}\cdot\frac{eV(s)}{E_0}\cdot\sin\left[h\cdot\frac{2\pi}{L}
\cdot\sigma+\varphi\right] \nonumber
\end{eqnarray}

Then we use series expansion of function $f(p_\sigma)$ from [7]:
\begin{equation}
f(p_\sigma)=f(0)+f^\prime(0)p_\sigma+f^{\prime\prime}(0)
\frac{1}{2}p_\sigma^2+\ldots=p_\sigma-\frac{1}{\gamma_0^2}
\cdot\frac{1}{2}p_\sigma^2+\ldots
\end{equation}
and the corresponding expansion of RHS of equations (4).
In the following we take into account only an arbitrary
polynomial (in terms of dynamical variables) expressions and
neglecting all nonpolynomial types of expressions, i.e. we
consider such approximations of RHS, which are not more than polynomial
functions in dynamical variables and arbitrary functions of
independent variable $s$ ("time" in our case, if we consider
our system of equations as dynamical problem).

\subsection{Particle in the Multipolar Field}

The magnetic vector potential of a magnet with $2n$ poles in Cartesian
coordinates is
\begin{equation}
A=\sum_n K_nf_n(x,y),
\end{equation}
where $ f_n$ is a homogeneous function of  $x$ and $y$ of order $n$.

The real and imaginary parts of binomial expansion of
\begin{equation}
f_n(x,y)=(x+iy)^n
\end{equation}
correspond to regular and skew multipoles. The cases $n=2$ to $n=5$
correspond to low-order multipoles: quadrupole, sextupole, octupole, decapole.

Then we have in particular case the following equations of motion for
single particle in a circular magnetic lattice in the transverse plane
$(x,y)$ ([8] for designation):
\begin{eqnarray}
\frac{\ud^2x}{\ud s^2}+ \left(\frac{1}{\rho(s)^2}-k_1(s)\right)x&=&
{\cal R}e\left[\sum_{n\geq 2}\frac{k_n(s)+ij_n(s)}{n!}\cdot(x+iy)^n\right]\\
\frac{\ud^2y}{\ud s^2}+k_1(s)y&=&-{\cal J}m\left[\sum_{n\geq}
\frac{k_n(s)+ij_n(s)}{n!}\cdot(x+iy)^n\right]\nonumber
\end{eqnarray}
and the corresponding Hamiltonian:
\begin{eqnarray}\label{eq:ham}
H(x,p_x,y,p_y,s)&=&\frac{p_x^2+p_y^2}{2}+\left(\frac{1}{\rho(s)^2}-k_1(s)\right)
\cdot\frac{x^2}{2}+k_1(s)\frac{y^2}{2}\\
&-&{\cal R}e\left[\sum_{n\geq 2}
\frac{k_n(s)+ij_n(s)}{(n+1)!}\cdot(x+iy)^{(n+1)}\right]\nonumber
\end{eqnarray}
Then we may take into account arbitrary but finite number in expansion
of RHS of Hamiltonian (\ref{eq:ham}) and
from our point of view the corresponding Hamiltonian equations of motions are
also not more than nonlinear ordinary differential equations with polynomial
nonlinearities and variable coefficients.

\subsection{Effects of Insertion Devices on Beam Dynamics}

Assuming a sinusoidal field variation, we may consider according to [9]
the analytical treatment of the effects of insertion devices on beam dynamics.
One of the major detrimental aspects of the installation of insertion devices
is the resulting reduction of dynamic aperture. Introduction of non-linearities
leads to enhancement of  the amplitude-dependent tune shifts and distortion of
phase space. The nonlinear fields will produce significant effects at large
betatron amplitudes.

The components of the insertion device magnetic field used for the derivation of
equations of motion are as follows:
\begin{eqnarray}
B_x&=&\frac{k_x}{k_y}\cdot B_0 \sinh(k_xx)\sinh(k_yy)\cos(kz)\nonumber\\
B_y&=&B_0\cosh(k_xx)\cosh(k_yy)\cos(kz)\\
B_z&=&-\frac{k}{k_y}B_0\cosh(k_xx)\sinh(k_yy)\sin(kz),\nonumber
\end{eqnarray}
with $k_x^2+k_y^2=k^2=(2\pi/\lambda)^2$,
where $\lambda$ is  the period length of the insertion device, $B_0$ is
its magnetic field, $\rho$ is the radius of the curvature in the field $B_0$.
After a canonical transformation to change to
betatron variables, the Hamiltonian is averaged over the period of
the insertion device and hyperbolic functions are expanded to the fourth
order in $x$ and $y$ (or arbitrary order).

Then we have the following Hamiltonian:
\begin{eqnarray}
H&=&\frac{1}{2}[p_x^2+p_y^2]+\frac{1}{4k^2\rho^2}[k_x^2x^2+k_y^2y^2]\nonumber\\
 &+&\frac{1}{12k^2\rho^2}[k_x^4x^4+k_y^4y^4+3k_x^2k^2x^2y^2]\\
 &-&\frac{\sin(ks)}{2k\rho}[p_x(k_x^2x^2+k_y^2y^2)-2k_xp_yxy]\nonumber
\end{eqnarray}
We have in this case also nonlinear (polynomial with degree 3) dynamical system with variable
(periodic) coefficients. As related cases we may consider wiggler and undulator
magnets. We have in horizontal $x-s$ plane the following equations
\begin{eqnarray}
\ddot{x}&=&-\dot{s}\frac{e}{m\gamma}{B_z(s)}\\
\ddot{s}&=&\dot{x}\frac{e}{m\gamma}B_z(s),\nonumber
\end{eqnarray}
where magnetic field has periodic dependence on $s$ and hyperbolic on $z$.

\subsection{Spin-Orbital Motion}
Let us consider the system of equations for orbital motion
\begin{equation}
\frac{\ud q}{\ud t}=\frac{\partial H_{orb}}{\partial p}, \qquad
\frac{\ud p}{\ud t}=-\frac{\partial H_{orb}}{\partial q}
\end{equation}
and Thomas-BMT equation for classical spin vector (see [10] for designation)
\begin{equation}
\frac{\ud s}{\ud t}=w\times s
\end{equation}
Here,
\begin{eqnarray}
H_{orb}&=&c\sqrt{\pi^2+m_0c^2}+e\Phi,\\
w&=&-\frac{e}{m_0\gamma c}\left( (1+\gamma G)\vec B-
    \frac{G(\vec\pi\cdot\vec B)\vec\pi}{m_0^2c^2(1+\gamma)}
 -\frac{1}{m_0c}\left(G+\frac{1}{1+\gamma}\right)[\pi\times E]
\right),\nonumber
\end{eqnarray}
$q=(q_1,q_2,q_3), p=(p_1,p_2,p_3)$ are canonical position and momentum,
$s=(s_1,s_2,s_3)$ is the classical spin vector of length $\hbar/2$,
$\pi=(\pi_1,\pi_2,\pi_3)$ is kinetic momentum vector.
We may introduce in 9-dimensional phase space $z=(q,p,s)$ the Poisson brackets
\begin{equation}
\{f(z),g(z)\}=f_qg_p-f_pg_q+[f_s\times g_s]\cdot s
\end{equation}
and the corresponding Hamiltonian equations:
\begin{equation}
\frac{\ud z}{\ud t}=\{z,H\},
\end{equation}
with Hamiltonian
\begin{equation}
H=H_{orb}(q,p,t)+w(q,p,t)\cdot s.
\end{equation}
More explicitly we have
\begin{eqnarray}
\frac{\ud q}{\ud t}&=&\frac{\partial H_{orb}}{\partial p}+\frac{\partial(w\cdot
  s)}{\partial p}\nonumber\\
\frac{\ud p}{\ud t}&=&-\frac{\partial H_{orb}}{\partial q}-\frac{\partial(w\cdot
  s)}{\partial q}\\
\frac{\ud s}{\ud t}&=&[w\times s]\nonumber
\end{eqnarray}
We will consider this dynamical system also in our second paper in this volume
via invariant approach, based on consideration of Lie-Poison structures on
semidirect products of groups.

But from the point of view which we used in this paper we may consider the
similar approximations as in preceding examples and then we also arrive to
some type of polynomial dynamics.

\section{Polynomial Dynamics}

The first main part of our consideration is some variational approach
to this problem, which reduces initial problem to the problem of
solution of functional equations at the first stage and some
algebraical problems at the second stage.
We have the solution in a compactly
supported wavelet basis.
Multiresolution expansion is the second main part of our construction.
The solution is parameterized by solutions of two reduced algebraical
problems, one is nonlinear and the second is some linear
problem, which is obtained from one of the next wavelet
constructions: Fast Wavelet Transform (FWT), Stationary
Subdivision Schemes (SSS), the method of Connection
Coefficients (CC).

\subsection{ Variational Method}
Our problems may be formulated as the systems of ordinary differential
equations
\begin{eqnarray}
{\ud x_i}/{\ud t}=f_i(x_j,t), \quad (i,j=1,...,n)
\end{eqnarray}
with fixed initial conditions $x_i(0)$, where $f_i$ are not more
than polynomial functions of dynamical variables $x_j$
and  have arbitrary dependence of time. Because of time dilation
we can consider  only next time interval: $0\leq t\leq 1$.
 Let us consider a set of
functions
\begin{eqnarray}
 \Phi_i(t)=x_i{\ud y_i}/{\ud t}+f_iy_i
\end{eqnarray}
and a set of functionals
\begin{eqnarray}
F_i(x)=\int_0^1\Phi_i (t)dt-x_iy_i\mid^1_0,
\end{eqnarray}
where $y_i(t) (y_i(0)=0)$ are dual variables.
It is obvious that the initial system  and the system
\begin{equation}
F_i(x)=0
\end{equation}
are equivalent.
In  part VI
we consider more general approach, which is based on possibility taking into
account underlying symplectic structure and on more useful and flexible
analytical approach, related to bilinear structure of initial functional.

Now we consider formal expansions for $x_i, y_i$:
\begin{eqnarray}\label{eq:pol1}
x_i(t)=x_i(0)+\sum_k\lambda_i^k\varphi_k(t)\quad
y_j(t)=\sum_r \eta_j^r\varphi_r(t),
\end{eqnarray}
where because of initial conditions we need only $\varphi_k(0)=0$.
Then we have the following reduced algebraical system
of equations on the set of unknown coefficients $\lambda_i^k$ of
expansions (\ref{eq:pol1}):
\begin{eqnarray}\label{eq:pol2}
\sum_k\mu_{kr}\lambda^k_i-\gamma_i^r(\lambda_j)=0
\end{eqnarray}
Its coefficients are
\begin{eqnarray}
\mu_{kr}=\int_0^1\varphi_k'(t)\varphi_r(t){\rm d}t,\quad
\gamma_i^r=\int_0^1f_i(x_j,t)\varphi_r(t){\rm d}t.
\end{eqnarray}
Now, when we solve system (\ref{eq:pol2}) and determine
 unknown coefficients from formal expansion (\ref{eq:pol1}) we therefore
obtain the solution of our initial problem.
It should be noted if we consider only truncated expansion (\ref{eq:pol1}) with N terms
then we have from (\ref{eq:pol2}) the system of $N\times n$ algebraical equations and
the degree of this algebraical system coincides
 with degree of initial differential system.
So, we have the solution of the initial nonlinear
(polynomial) problem  in the form
\begin{eqnarray}\label{eq:pol3}
x_i(t)=x_i(0)+\sum_{k=1}^N\lambda_i^k X_k(t),
\end{eqnarray}
where coefficients $\lambda_i^k$ are roots of the corresponding
reduced algebraical problem (\ref{eq:pol2}).
Consequently, we have a parametrization of solution of initial problem
by solution of reduced algebraical problem (\ref{eq:pol2}).
The first main problem is a problem of
 computations of coefficients of reduced algebraical
system.
As we will see, these problems may be explicitly solved in wavelet approach.

Next we consider the  construction  of explicit time
solution for our problem. The obtained solutions are given
in the form (\ref{eq:pol3}),
where
$X_k(t)$ are basis functions and
  $\lambda_k^i$ are roots of reduced
 system of equations.  In our first wavelet case $X_k(t)$
are obtained via multiresolution expansions and represented by
 compactly supported wavelets and $\lambda_k^i$ are the roots of
corresponding general polynomial  system (\ref{eq:pol2})  with coefficients, which
are given by FWT, SSS or CC constructions.  According to the
        variational method   to  give the reduction from
differential to algebraical system of equations we need compute
the objects $\gamma ^j_a$ and $\mu_{ji}$,
which are constructed from objects:
\begin{eqnarray}\label{eq:pol4}
\sigma_i&\equiv&\int^1_0X_i(\tau)\ud\tau,\nonumber\\
   \nu_{ij}&\equiv&\int^1_0X_i(\tau)X_j(\tau)\ud\tau,\nonumber\\
    \mu_{ji}&\equiv&\int X'_i(\tau)X_j(\tau)\ud\tau,\\
   \beta_{klj}&\equiv&\int^1_0X_k(\tau)X_l(\tau)X_j(\tau)\ud\tau \nonumber
\end{eqnarray}
for the simplest case of Riccati systems, where degree of nonlinearity equals to
two. For the general case of arbitrary n we have analogous to (\ref{eq:pol4})
iterated integrals with the degree of monomials in integrand which is one more
bigger than degree of initial system.

\subsection{Wavelet Framework}
Our constructions are based on multiresolution approach. Because affine
group of translation and dilations is inside the approach, this
method resembles the action of a microscope. We have contribution to
final result from each scale of resolution from the whole
infinite scale of spaces. More exactly, the closed subspace
$V_j (j\in {\bf Z})$ corresponds to  level j of resolution, or to scale j.
We consider  a r-regular multiresolution analysis of $L^2 ({\bf R}^n)$
(of course, we may consider any different functional space)
which is a sequence of increasing closed subspaces $V_j$:
\begin{equation}
...V_{-2}\subset V_{-1}\subset V_0\subset V_{1}\subset V_{2}\subset ...
\end{equation}
satisfying the following properties:
\begin{eqnarray}
&&\displaystyle\bigcap_{j\in{\bf Z}}V_j=0,\quad
\overline{\displaystyle\bigcup_{j\in{\bf Z}}}V_j=L^2({\bf R}^n),\nonumber\\
&& f(x)\in V_j <=> f(2x)\in V_{j+1}, \nonumber\\
&& f(x)\in V_0 <=> f(x-k)\in V_0, \quad, \forall k\in {\bf Z}^n.
\end{eqnarray}
There exists a function $\varphi\in V_0$ such that
\{${\varphi_{0,k}(x)=
\varphi(x-k)}, k\in{\bf Z}^n$\} forms a Riesz basis for $V_0$.

The function $\varphi$ is regular and localized:
$\varphi$ is $C^{r-1}, \quad \varphi^{(r-1)}$ is almost
everywhere differentiable and for almost every $x\in {\bf R}^n$, for
every integer $\alpha\leq r$ and for all integer p there exists
constant $C_p$ such that
\begin{equation}
\mid\partial^\alpha \varphi(x)\mid \leq C_p(1+|x|)^{-p}
\end{equation}

Let
 $\varphi(x)$ be
a scaling function, $\psi(x)$ is a wavelet function and
$\varphi_i(x)=\varphi(x-i)$. Scaling relations that define
$\varphi,\psi$ are
\begin{eqnarray}
\varphi(x)&=&\sum\limits^{N-1}_{k=0}a_k\varphi(2x-k)=
\sum\limits^{N-1}_{k=0}a_k\varphi_k(2x),\\
\psi(x)&=&\sum\limits^{N-2}_{k=-1}(-1)^k a_{k+1}\varphi(2x+k).
\end{eqnarray}
Let  indices $\ell, j$
 represent translation and scaling, respectively and
\begin{equation}
\varphi_{jl}(x)=2^{j/2}\varphi(2^j x-\ell)
\end{equation}
then the set $\{\varphi_{j,k}\}, {k\in {\bf Z}^n}$ forms a Riesz basis for $V_j$.
The wavelet function $\psi $ is used to encode the details between
two successive levels of approximation.
Let $W_j$ be the orthonormal complement of $V_j$ with respect to $V_{j+1}$:
\begin{equation}
V_{j+1}=V_j\bigoplus W_j.
\end{equation}
Then just as $V_j$ is spanned by dilation and translations of the scaling function,
so are $W_j$ spanned by translations and dilation of the mother wavelet
$\psi_{jk}(x)$, where
\begin{equation}
\psi_{jk}(x)=2^{j/2}\psi(2^j x-k).
\end{equation}
All expansions which we used are based on the following properties:
\begin{eqnarray}
&&\{\psi_{jk}\}, \quad j,k\in {\bf Z}\quad
  \mbox{is a Hilbertian basis of } L^2({\bf R})\nonumber\\
&&\{\varphi_{jk}\}_{j\geq 0, k\in {\bf Z}} \quad\mbox{is an orthonormal
basis for} L^2({\bf R}),\nonumber\\
&& L^2({\bf R})=\overline{V_0\displaystyle\bigoplus^\infty_{j=0} W_j},\\
&& \mbox{or}\nonumber\\
&&\{\varphi_{0,k},\psi_{j,k}\}_{j\geq 0,k\in {\bf Z}} \quad\mbox{is
an orthonormal basis for}
 L^2({\bf R}).\nonumber
\end{eqnarray}

\subsection{ Wavelet Computations}
 Now we give construction for
computations of objects (28) in the wavelet case.
We use compactly supported wavelet basis: orthonormal basis
for functions in $L^2({\bf R})$.

Let be  $ f : {\bf R}\longrightarrow {\bf C}$ and the wavelet
expansion is
\begin{eqnarray}
f(x)=\sum\limits_{\ell\in{\bf Z}}c_\ell\varphi_\ell(x)+
\sum\limits_{j=0}^\infty\sum\limits_{k\in{\bf
Z}}c_{jk}\psi_{jk}(x)
\end{eqnarray}

If in formulae (38) $c_{jk}=0$ for $j\geq J$, then $f(x)$ has an alternative
expansion in terms of dilated scaling functions only
$
f(x)=\sum\limits_{\ell\in {\bf Z}}c_{J\ell}\varphi_{J\ell}(x)
$.
This is a finite wavelet expansion, it can be written solely in
terms of translated scaling functions.
Also we have the shortest possible support: scaling function
$DN$ (where $N$ is even integer) will have support $[0,N-1]$ and
$N/2$ vanishing moments.
There exists $\lambda>0$ such that $DN$ has $\lambda N$
continuous derivatives; for small $N,\lambda\geq 0.55$.
To solve our second associated linear problem we need to
evaluate derivatives of $f(x)$ in terms of $\varphi(x)$.
Let be $
\varphi^n_\ell=\ud^n\varphi_\ell(x)/\ud x^n
$.
We consider computation of the wavelet - Galerkin integrals.
Let $f^d(x)$ be d-derivative of function
 $f(x)$, then we have
$
f^d(x)=\sum_\ell c_l\varphi_\ell^d(x)
$,
and values $\varphi_\ell^d(x)$ can be expanded in terms of
$\varphi(x)$
\begin{eqnarray}
\phi_\ell^d(x)&=&\sum\limits_m\lambda_m\varphi_m(x),\\
\lambda_m&=&\int\limits_{-\infty}^{\infty}\varphi_\ell^d(x)\varphi_m(x)\ud x,\nonumber
 \end{eqnarray}
where $\lambda_m$ are wavelet-Galerkin integrals.
The coefficients $\lambda_m$  are 2-term connection
coefficients. In general we need to find $(d_i\geq 0)$
\begin{eqnarray}
\Lambda^{d_1 d_2 ...d_n}_{\ell_1 \ell_2 ...\ell_n}=
 \int\limits_{-\infty}^{\infty}\prod\varphi^{d_i}_{\ell_i}(x)dx
\end{eqnarray}
For Riccati case we need to evaluate two and three
connection coefficients
\begin{eqnarray}
\Lambda_\ell^{d_1
d_2}=\int^\infty_{-\infty}\varphi^{d_1}(x)\varphi_\ell^{d_2}(x)dx,
\quad
\Lambda^{d_1 d_2
d_3}=\int\limits_{-\infty}^\infty\varphi^{d_1}(x)\varphi_
\ell^{d_2}(x)\varphi^{d_3}_m(x)dx
\end{eqnarray}
According to CC method [11] we use the next construction. When $N$  in
scaling equation is a finite even positive integer the function
$\varphi(x)$  has compact support contained in $[0,N-1]$.
For a fixed triple $(d_1,d_2,d_3)$ only some  $\Lambda_{\ell
 m}^{d_1 d_2 d_3}$ are nonzero: $2-N\leq \ell\leq N-2,\quad
2-N\leq m\leq N-2,\quad |\ell-m|\leq N-2$. There are
$M=3N^2-9N+7$ such pairs $(\ell,m)$. Let $\Lambda^{d_1 d_2 d_3}$
be an M-vector, whose components are numbers $\Lambda^{d_1 d_2
d_3}_{\ell m}$. Then we have the first reduced algebraical system
: $\Lambda$
satisfy the system of equations $(d=d_1+d_2+d_3)$
\begin{eqnarray}
A\Lambda^{d_1 d_2 d_3}=2^{1-d}\Lambda^{d_1 d_2 d_3},
\qquad
A_{\ell,m;q,r}=\sum\limits_p a_p a_{q-2\ell+p}a_{r-2m+p}
\end{eqnarray}
By moment equations we have created a system of $M+d+1$
equations in $M$ unknowns. It has rank $M$ and we can obtain
unique solution by combination of LU decomposition and QR
algorithm.
The second  reduced algebraical system gives us the 2-term connection
coefficients:
\begin{eqnarray}
A\Lambda^{d_1 d_2}=2^{1-d}\Lambda^{d_1 d_2},\quad d=d_1+d_2,\quad
A_{\ell,q}=\sum\limits_p a_p a_{q-2\ell+p}
\end{eqnarray}
For nonquadratic case we have analogously additional linear problems for
objects (40).
Solving these linear problems we obtain the coefficients of nonlinear
algebraical system (25) and after that we obtain the coefficients of wavelet
expansion (27). As a result we obtained the explicit time solution  of our
problem in the base of compactly supported wavelets.
We use for modelling D6, D8, D10 functions and programs RADAU and
DOPRI for testing.

In the following we consider extension of this approach to the case of periodic
boundary conditions, the case of presence of arbitrary variable coefficients
and more flexible biorthogonal wavelet approach.

\section{Evaluation of Nonlinearities Scale by Scale}
\subsection{Para-product and Decoupling between Scales}
But before we consider two different schemes of modification of our variational
approach in the case when we consider different scales separately. For this reason
we need to compute errors of approximations. The main problems come of course
from nonlinear terms. We follow approach from [12].

Let $P_j$ be projection operators on the subspaces $V_j, j\in{\bf Z}$:
\begin{eqnarray}
P_j &:& L^2({\bf R}) \to V_j\\
(P_j f)(x)&=&\sum_k <f,\varphi_{j,k}> \varphi_{j,k}(x)\nonumber
\end{eqnarray}
and $Q_j$ are projection operators on the subspaces $W_j$:
\begin{eqnarray}
Q_j=P_{j-1}-P_j
\end{eqnarray}
So, for $u\in L^2({\bf R})$ we have $u_j=P_ju\quad$ and $u_j\in V_j$,
where $\{V_j\}, j\in{\bf Z}$ is a multiresolution analysis of $L^2({\bf R})$.
It is obviously that we can represent $u_0^2$ in the following form:
\begin{equation}\label{eq:form1}
u_0^2=2\sum^n_{j=1}(P_ju)(Q_ju)+\sum^n_{j=1}(Q_ju)(Q_ju)+u_n^2
\end{equation}
In this formula there is no interaction between different scales.
We may consider each term of (\ref{eq:form1}) as a bilinear mappings:
\begin{eqnarray}\label{eq:form2}
\displaystyle
M_{VW}^j : V_j\times W_j\to L^2({\bf R})=
V_j{\oplus_{j'\geq j}W_{j'}}
\end{eqnarray}
\begin{eqnarray}\label{eq:form3}
M_{WW}^j : W_j\times W_j\to L^2({\bf R})=V_j\oplus_{j'\geq j}W_{j'}
\end{eqnarray}
For numerical purposes we need formula (\ref{eq:form1}) with a finite number of
scales, but when we consider limits $j\to\infty$ we have
\begin{equation}
u^2=\sum_{j\in {\bf Z}}(2P_ju+Q_ju)(Q_ju),
\end{equation}
which is para-product of Bony, Coifman and Meyer.

Now we need to expand (\ref{eq:form1}) into the wavelet bases. To expand
each term in (\ref{eq:form1}) into wavelet basis, we need to consider
the integrals of the products of the basis functions, e.g.
\begin{equation}
M^{j,j'}_{WWW}(k,k',\ell)=\int^\infty_{-\infty}\psi^j_k(x)
\psi^j_{k'}(x)\psi^{j'}_\ell(x)\ud x,
\end{equation}
where $j'>j$ and
\begin{equation}
\psi^j_k(x)=2^{-j/2}\psi(2^{-j}x-k)
\end{equation}
are the basis functions.
If we consider compactly supported wavelets then
\begin{equation}
M_{WWW}^{j,j'}(k,k',\ell)\equiv 0\quad \mbox{for}\quad |k-k'|>k_0,
\end{equation}
where $k_0$ depends on the overlap of the supports of the basis functions
and
\begin{equation}\label{eq:form4}
|M_{WWW}^r(k-k',2^rk-\ell)|\leq C\cdot 2^{-r\lambda M}
\end{equation}
Let us define $j_0$ as the distance between scales such that for a given
$\varepsilon$ all the coefficients in (\ref{eq:form4}) with labels
$r=j-j'$, $r>j_0$ have absolute values less than $\varepsilon$. For the purposes of
computing with accuracy $\varepsilon$ we replace the mappings in
(\ref{eq:form2}), (\ref{eq:form3}) by
\begin{equation}\label{eq:z1}
M_{VW}^j : V_j\times W_j\to V_j\oplus_{j\leq j'\leq j_0}W_{j'}
\end{equation}
\begin{equation}\label{eq:z2}
M_{WW}^j : W_j\times W_j\to V_j\oplus_{J\leq j'\leq j_0}W_{j'}\nonumber
\end{equation}
Since
\begin{equation}
V_j\oplus_{j\leq j'\leq j_0}W_{j'}=V_{j_0-1}
\end{equation}
and
\begin{equation}
V_j\subset V_{j_0-1},\qquad W_j\subset V_{j_0-1}
\end{equation}
we may consider bilinear mappings (\ref{eq:z1}), (\ref{eq:z2}) on
$V_{j_0-1}\times V_{j_0-1}$.
For the evaluation of (\ref{eq:z1}), (\ref{eq:z2}) as mappings
$V_{j_0-1}\times V_{j_0-1}\to V_{j_0-1}$
we need significantly fewer coefficients than for
mappings (\ref{eq:z1}), (\ref{eq:z2}). It is enough to consider only
coefficients
\begin{equation}
M(k,k',\ell)=2^{-j/2}\int^\infty_\infty\varphi(x-k)\varphi(x-k')\varphi(x-\ell)\ud
x,
\end{equation}
where $\varphi(x)$ is scale function. Also we have
\begin{equation}
M(k,k',\ell)=2^{-j/2}M_0(k-\ell,k'-\ell),
\end{equation}
where
\begin{equation}
M_0(p,q)=\int\varphi(x-p)\varphi(x-q)\varphi(x)\ud x
\end{equation}
Now as in section (3C) we may derive and solve a system of linear equations to
find $M_0(p,q)$.

\subsection{Non-regular Approximation}

We use wavelet function $\psi(x)
$, which has $k$ vanishing moments
$
\int x^k \psi(x)\ud x=0$, or equivalently
$x^k=\sum c_\ell\varphi_\ell(x)$ for each $k$,
$0\leq k\leq K$.

Let $P_j$ again be orthogonal projector on space $V_j$. By tree algorithm we
have for any $u\in L^2({\bf R})$ and $\ell\in{\bf Z}$, that the wavelet
coefficients of $P_\ell(u)$, i.e. the set $\{<u,\psi_{j,k}>, j\leq\ell-1,
k\in{\bf Z}\}$ can be compute using hierarchic algorithms from the set of
scaling coefficients in $V_\ell$, i.e. the set $\{<u,\varphi_{\ell,k}>,
k\in{\bf Z}\}$ [13]. Because for scaling function $\varphi$ we have in general
only $\int\varphi(x)\ud x=1$, therefore we have for any function $u\in L^2({\bf
R})$:
\begin{equation}
\lim_{j\to\infty, k2^{-j}\to x} \mid 2^{j/2}<u,\varphi_{j,k}>-u(x)\mid=0
\end{equation}
If the integer $n(\varphi)$ is the largest one such that
\begin{equation}
\int x^\alpha \varphi(x)\ud x=0 \qquad \mbox{for}\qquad 1\leq\alpha\leq n
\end{equation}
then if $u\in C^{(n+1)}$ with $u^{(n+1)}$ bounded we have for $j\to\infty$
uniformly in k:
\begin{equation}
\mid 2^{j/2}<u,\varphi_{j,k}>-u(k2^{-j})\mid=O(2^{-j(n+1)}).
\end{equation}
Such scaling functions with zero moments are very useful for us from the point
of view of time-frequency localization, because we have for Fourier component
$\hat\Phi(\omega)$ of them, that exists some $C(\varphi)\in{\bf R}$, such
that for $\omega\to0$ $\quad\hat\Phi(\omega)=1+C(\varphi)$ $\mid\omega\mid^{2r+2}$
(remember, that we consider r-regular multiresolution analysis).
Using such type of scaling functions lead to superconvergence properties for
general Galerkin approximation [13].
Now we need some estimates in each scale for non-linear terms of type $u\mapsto
f(u)=f\circ u$, where f is $C^\infty$ (in previous and future parts we consider
only truncated Taylor series action). Let us consider non regular space of
approximation $\widetilde V$ of the form
\begin{equation}\label{eq:til1}
\widetilde V=V_q\oplus\sum_{q\leq j\leq p-1} \widetilde{W_j},
\end{equation}
with $\widetilde{W_j}\subset W_j$. We need efficient and precise estimate of
$f\circ u$ on $\widetilde V$. Let us set for $q\in{\bf Z}$ and $u\in L^2({\bf
R})$
\begin{equation}
\prod f_q (u)=2^{-q/2}\sum_{k\in{\bf
Z}}f(2^{q/2}<u,\varphi_{q,k}>)\cdot\varphi_{q,k}
\end{equation}
We have the following important for us estimation (uniformly in q) for $u,
f(u)\in H^{(n+1)}$ [13]:
\begin{equation}\label{eq:til2}
\|P_q\left(f(u)\right)-\prod f_q(u)\|_{L^2}=O\left(2^{-(n+1)q}\right)
\end{equation}
For non regular spaces (\ref{eq:til1}) we set
\begin{equation}
\prod f_{\widetilde V}(u)=\prod f_q(u)+\sum_{\ell=q,p-1} P_{\widetilde {W_j}}
\prod f_{\ell+1}(u)
\end{equation}
Then we have the following estimate:
\begin{equation}
\Vert P_{\widetilde V}\left(f(u)\right)-\prod f_{\widetilde V}(u)\Vert_{L^2}=O(2^{-(n+1)q})
\end{equation}
uniformly in q and $\widetilde V$ (\ref{eq:til1}).

This estimate depends on q, not p, i.e. on the scale of the coarse grid, not on
the finest grid used in definition of $\widetilde V$. We have for total error
\begin{equation}
\Vert f(u)-\prod f_{\widetilde V}(u)\Vert=
\Vert f(u)-P_{\widetilde V}(f(u))\Vert_{L^2}+
\Vert P_{\widetilde V}(f(u)-\prod f_{\widetilde V}(u))\Vert_{L^2}
\end{equation}
and since the projection error in $\widetilde V$:
$
\Vert f(u)-P_{\bar{V}}\left(f(u)\right)\Vert_{L^2}
$
 is much smaller than the projection error in
$V_q$ we have the improvement (68) of (\ref{eq:til2}).
In our concrete calculations and estimates it is very useful to consider
our approximations in the particular case of
c-structured space:
\begin{equation}
\widetilde{V}=V_q+\sum^{p-1}_{j=q}span\{\psi_{j,k}, k\in[2^{(j-1)}-c,
2^{(j-1)}+c]
\quad\mbox{\rm mod}\quad 2^j\}
\end{equation}

\section{VARIATIONAL WAVELET APPROACH\\ FOR PERIODIC TRAJECTORIES}

We start with extension of our approach to the case
of periodic trajectories. The equations of motion corresponding
to Hamiltonians (from part II) may also be formulated as a particular case of
the general system of
ordinary differential equations
$
{dx_i}/{dt}=f_i(x_j,t)$, $  (i,j=1,...,n)$, $0\leq t\leq 1$,
where $f_i$ are not more
than polynomial functions of dynamical variables $x_j$
and  have arbitrary dependence of time but with periodic boundary conditions.
According to our variational approach we have the
solution in the following form
\begin{eqnarray}
x_i(t)=x_i(0)+\sum_k\lambda_i^k\varphi_k(t),\qquad x_i(0)=x_i(1),
\end{eqnarray}
where $\lambda_i^k$ are again the roots of reduced algebraical
systems of equations
with the same degree of nonlinearity and $\varphi_k(t)$
corresponds to useful type of wavelet bases (frames).
It should be noted that coefficients of reduced algebraical system
are the solutions of additional linear problem and
also
depend on particular type of wavelet construction and type of bases.

This linear problem is our second reduced algebraical problem. We need to find
in general situation objects
\begin{eqnarray}
\Lambda^{d_1 d_2 ...d_n}_{\ell_1 \ell_2 ...\ell_n}=
 \int\limits_{-\infty}^{\infty}\prod\varphi^{d_i}_{\ell_i}(x)\ud x,
\end{eqnarray}
but now in the case of periodic boundary conditions.
Now we consider the procedure of their
calculations in case of periodic boundary conditions
 in the base of periodic wavelet functions on
the interval [0,1] and corresponding expansion (71) inside our
variational approach. Periodization procedure
gives us
\begin{eqnarray}
\hat\varphi_{j,k}(x)&\equiv&\sum_{\ell\in Z}\varphi_{j,k}(x-\ell)\\
\hat\psi_{j,k}(x)&=&\sum_{\ell\in Z}\psi_{j,k}(x-\ell)\nonumber
\end{eqnarray}
So, $\hat\varphi, \hat\psi$ are periodic functions on the interval
 [0,1]. Because $\varphi_{j,k}=\varphi_{j,k'}$ if $k=k'\mathrm{mod}(2^j)$, we
may consider only $0\leq k\leq 2^j$ and as  consequence our
multiresolution has the form
$\displaystyle\bigcup_{j\geq 0} \hat V_j=L^2[0,1]$ with
$\hat V_j= \mathrm{span} \{\hat\varphi_{j,k}\}^{2j-1}_{k=0}$ [14].
Integration by parts and periodicity gives  useful relations between
objects (72) in particular quadratic case $(d=d_1+d_2)$:
\begin{eqnarray}
\Lambda^{d_1,d_2}_{k_1,k_2}=(-1)^{d_1}\Lambda^{0,d_2+d_1}_{k_1,k_2},
\Lambda^{0,d}_{k_1,k_2}=\Lambda^{0,d}_{0,k_2-k_1}\equiv
\Lambda^d_{k_2-k_1}
\end{eqnarray}
So, any 2-tuple can be represent by $\Lambda^d_k$.
Then our second additional linear problem is reduced to the eigenvalue
problem for
$\{\Lambda^d_k\}_{0\leq k\le 2^j}$ by creating a system of $2^j$
homogeneous relations in $\Lambda^d_k$ and inhomogeneous equations.
So, if we have dilation equation in the form
$\varphi(x)=\sqrt{2}\sum_{k\in Z}h_k\varphi(2x-k)$,
then we have the following homogeneous relations
\begin{equation}
\Lambda^d_k=2^d\sum_{m=0}^{N-1}\sum_{\ell=0}^{N-1}h_m h_\ell
\Lambda^d_{\ell+2k-m},
\end{equation}
or in such form
$A\lambda^d=2^d\lambda^d$, where $\lambda^d=\{\Lambda^d_k\}_
{0\leq k\le 2^j}$.
Inhomogeneous equations are:
\begin{equation}
\sum_{\ell}M_\ell^d\Lambda^d_\ell=d!2^{-j/2},
\end{equation}
 where objects
$M_\ell^d(|\ell|\leq N-2)$ can be computed by recursive procedure
\begin{equation}
M_\ell^d=2^{-j(2d+1)/2}\tilde{M_\ell^d}, \quad\tilde{M_\ell^k}=
<x^k,\varphi_{0,\ell}>=\sum_{j=0}^k {k\choose j} n^{k-j}M_0^j,\quad
\tilde{M_0^\ell}=1.
\end{equation}
 So, we reduced our last problem to standard
linear algebraical problem. Then we use the same methods as in part III C.
As a result we obtained
for closed trajectories of orbital dynamics described by Hamiltonians from part
II
the explicit time solution (71) in the base of periodized wavelets (73).

\section{VARIATIONAL APPROACH IN BIORTHO\-GONAL WAVELET BASES}

Now we consider further generalization of our variational wavelet approach.
In [1]-[3] we consider different types of variational principles
which give us weak solutions of our nonlinear problems.

Before we consider the generalization of our wavelet variational
approach  to the symplectic invariant calculation of
closed loops in Hamiltonian systems
[3]. We also have the parametrization of our solution by some
reduced algebraical problem but in contrast to the general case where
the solution is parametrized by construction based on scalar
refinement equation, in symplectic case we have
parametrization of the solution
by matrix problems -- Quadratic Mirror Filters equations [3].
But because integrand of variational functionals is represented
by bilinear form (scalar product) it seems more reasonable to
consider wavelet constructions [15] which take into account all advantages of
this structure.

The action functional for loops in the phase space is [16]
\begin{equation}
F(\gamma)=\displaystyle\int_\gamma pdq-\int_0^1H(t,\gamma(t))dt
\end{equation}
The critical points of $F$ are those loops $\gamma$, which solve
the Hamiltonian equations associated with the Hamiltonian $H$
and hence are periodic orbits. By the way, all critical points of $F$ are
the saddle points of infinite Morse index, but surprisingly this approach  is
very effective. This will be demonstrated using several
variational techniques starting from minimax due to Rabinowitz
and ending with Floer homology. So, $(M,\omega)$ is symplectic
manifolds, $H: M \to R $, $H$ is Hamiltonian, $X_H$ is
unique Hamiltonian vector field defined  by
\begin{equation}
\omega(X_H(x),\upsilon)=-dH(x)(\upsilon),\quad \upsilon\in T_xM,
\quad x\in M,
\end{equation}
where $ \omega$ is the symplectic structure.
A T-periodic solution $x(t)$ of the Hamiltonian equations
\begin{equation}
\dot x=X_H(x) \quad \mbox{ on $M$}
\end{equation}
is a solution, satisfying the boundary conditions $x(T)$ $=x(0), T>0$.
Let us consider the loop space $\Omega=C^\infty(S^1, R^{2n})$,
where $S^1=R/{\bf Z}$, of smooth loops in $R^{2n}$.
Let us define a function $\Phi: \Omega\to R $ by setting
\begin{equation}
\Phi(x)=\displaystyle\int_0^1\frac{1}{2}<-J\dot x, x>dt-
\int_0^1 H(x(t))dt, \quad x\in\Omega
\end{equation}
The critical points of $\Phi$ are the periodic solutions of $\dot x=X_H(x)$.
Computing the derivative at $x\in\Omega$ in the direction of $y\in\Omega$,
we find
\begin{eqnarray}
\Phi'(x)(y)=\frac{d}{d\epsilon}\Phi(x+\epsilon y)\vert_{\epsilon=0}
=
\displaystyle\int_0^1<-J\dot x-\bigtriangledown H(x),y>dt
\end{eqnarray}
Consequently, $\Phi'(x)(y)=0$ for all $y\in\Omega$ iff the loop $x$ satisfies
the equation
\begin{equation}
-J\dot x(t)-\bigtriangledown H(x(t))=0,
\end{equation}
i.e. $x(t)$ is a solution of the Hamiltonian equations, which also satisfies
$x(0)=x(1)$, i.e. periodic of period 1. Periodic loops may be represented by
their Fourier series:
\begin{equation}
x(t)=\displaystyle\sum_{k\in{\bf Z}}e^{k2\pi Jt}x_k, \quad x_k\in R^{2k},
\end{equation}
where $J$ is quasicomplex structure. We give relations between
quasicomplex structure and wavelets in our second paper in this volume
(see also [3]).
But now we
need to take into account underlying bilinear structure via wavelets.

We started with two hierarchical sequences of approximations spaces [15]:
\begin{eqnarray}
\dots V_{-2}\subset V_{-1}\subset V_{0}\subset V_{1}\subset V_{2}\dots,\qquad
\dots \widetilde{V}_{-2}\subset\widetilde{V}_{-1}\subset
\widetilde{V}_{0}\subset\widetilde{V}_{1}\subset\widetilde{V}_{2}\dots,
\end{eqnarray}
and as usually,
$W_0$ is complement to $V_0$ in $V_1$, but now not necessarily orthogonal
complement.
New orthogonality conditions have now the following form:
\begin{equation}
\widetilde {W}_{0}\perp V_0,\qquad  W_{0}\perp\widetilde{V}_{0},\qquad
V_j\perp\widetilde{W}_j, \qquad \widetilde{V}_j\perp W_j
\end{equation}
translates of $\psi$ $\mathrm{span}$ $ W_0$,
translates of $\tilde\psi \quad \mathrm{span} \quad\widetilde{W}_0$.
Biorthogonality conditions are
\begin{equation}
<\psi_{jk},\tilde{\psi}_{j'k'}>=
\int^\infty_{-\infty}\psi_{jk}(x)\tilde\psi_{j'k'}(x)\ud x=
\delta_{kk'}\delta_{jj'},
\end{equation}
 where
$\psi_{jk}(x)=2^{j/2}\psi(2^jx-k)$.
Functions $\varphi(x), \tilde\varphi(x-k)$ form  dual pair:
\begin{equation}
<\varphi(x-k),\tilde\varphi(x-\ell)>=\delta_{kl},\quad
 <\varphi(x-k),\tilde\psi(x-\ell)>=0\quad  \mbox{for}\quad \forall k,
\ \forall\ell.
\end{equation}
Functions $\varphi, \tilde\varphi$ generate a multiresolution analysis.
$\varphi(x-k)$, $\psi(x-k)$ are synthesis functions,
$\tilde\varphi(x-\ell)$, $\tilde\psi(x-\ell)$ are analysis functions.
Synthesis functions are biorthogonal to analysis functions. Scaling spaces
are orthogonal to dual wavelet spaces.
Two multiresolutions are intertwining
$
V_j+W_j=V_{j+1}, \quad \widetilde V_j+ \widetilde W_j = \widetilde V_{j+1}
$.
These are direct sums but not orthogonal sums.

So, our representation for solution has now the form
\begin{equation}
f(t)=\sum_{j,k}\tilde b_{jk}\psi_{jk}(t),
\end{equation}
where synthesis wavelets are used to synthesize the function. But
$\tilde b_{jk}$ come from inner products with analysis wavelets.
Biorthogonality yields
\begin{equation}
\tilde b_{\ell m}=\int f(t)\tilde{\psi}_{\ell m}(t) \ud t.
\end{equation}
So, now we can introduce this more complicated construction into
our variational approach. We have modification only on the level of
computing coefficients of reduced nonlinear algebraical system.
This new construction is more flexible.
Biorthogonal point of view is more stable under the action of large
class of operators while orthogonal (one scale for multiresolution)
is fragile, all computations are much more simpler and we accelerate
the rate of convergence. In all types of Hamiltonian calculation,
which are based on some bilinear structures (symplectic or
Poissonian structures, bilinear form of integrand in variational
integral) this framework leads to greater success.

\section{Variable Coefficients}
In the case when we have situation when our problem is described a system of
nonlinear (polynomial)differential equations, we need to consider
extension of our previous approach which can take into  account
any type of variable coefficients (periodic, regular or singular).
We can produce such approach if we add in our construction additional
refinement equation, which encoded all information about variable
coefficients [17].
According to our variational approach we need to compute integrals of
the form
\begin{equation}\label{eq:var1}
\int_Db_{ij}(t)(\varphi_1)^{d_1}(2^m t-k_1)(\varphi_2)^{d_2}
(2^m t-k_2)\ud x,
\end{equation}
where now $b_{ij}(t)$ are arbitrary functions of time, where trial
functions $\varphi_1,\varphi_2$ satisfy a refinement equations:
\begin{equation}
\varphi_i(t)=\sum_{k\in{\bf Z}}a_{ik}\varphi_i(2t-k)
\end{equation}
If we consider all computations in the class of compactly supported wavelets
then only a finite number of coefficients do not vanish. To approximate
the non-constant coefficients, we need choose a different refinable function
$\varphi_3$ along with some local approximation scheme
\begin{equation}
(B_\ell f)(x):=\sum_{\alpha\in{\bf Z}}F_{\ell,k}(f)\varphi_3(2^\ell t-k),
\end{equation}
where $F_{\ell,k}$ are suitable functionals supported in a small neighborhood
of $2^{-\ell}k$ and then replace $b_{ij}$ in (\ref{eq:var1}) by
$B_\ell b_{ij}(t)$. In particular case one can take a characteristic function
and can thus approximate non-smooth coefficients locally. To guarantee
sufficient accuracy of the resulting approximation to (\ref{eq:var1})
it is important to have the flexibility of choosing $\varphi_3$ different
from $\varphi_1, \varphi_2$. In the case when D is some domain, we
can write
\begin{equation}
b_{ij}(t)\mid_D=\sum_{0\leq k\leq 2^\ell}b_{ij}(t)\chi_D(2^\ell t-k),
\end{equation}
where $\chi_D$ is characteristic function of D. So, if we take
$\varphi_4=\chi_D$, which is again a refinable function, then the problem of
computation of (\ref{eq:var1}) is reduced to the problem of calculation of
integral
\begin{eqnarray}
&&H(k_1,k_2,k_3,k_4)=H(k)=\\
&&\int_{{\bf R}^s}\varphi_4(2^j t-k_1)\varphi_3(2^\ell t-k_2)
\varphi_1^{d_1}(2^r t-k_3)
\varphi_2^{d_2}(2^st-k_4)\ud x\nonumber
\end{eqnarray}
The key point is that these integrals also satisfy some sort of refinement
equation:
\begin{equation}
2^{-|\mu|}H(k)=\sum_{\ell\in{\bf Z}}b_{2k-\ell}H(\ell),\qquad \mu=d_1+d_2.
\end{equation}
This equation can be interpreted as the problem of computing an eigenvector.
Thus, we reduced the problem of extension of our method to the case of
variable coefficients to the same standard algebraical problem as in
the preceding sections. So, the general scheme is the same one and we
have only one more additional
linear algebraic problem by which we in the same way can parameterize the
solutions of corresponding problem.

\vspace{5mm}
Extended version and related results may be found in [1]-[6].

\section*{ACKNOWLEDGEMENTS}
We would like to thank Professors M.~Cornacchia, C.~Pellegrini,
L.~Palumbo, Mrs. M.~Laraneta, J.~Kono, G.~Nanula for
nice hospitality, help and support before and during Arcidosso meeting
and all participants for interesting discussions.
We are very grateful to Professor M.~Cornacchia, Mrs. J.~Kono and
M.~Laraneta, because without their permanent encouragement this paper
would not have been written.

 \end{document}